\documentclass[a4paper]{spie}  

 \usepackage{todonotes}
 
\usepackage{amsmath,amsfonts,amssymb}
\usepackage{graphicx}
\usepackage{cite}
\usepackage{enumitem}

\usepackage[colorlinks=true, allcolors=blue]{hyperref}

\title{Deep Learning for Improved Polyp Detection from Synthetic Narrow-Band Imaging}

\author[a]{Mathias Ramm Haugland}
\author[b]{Hemin Ali Qadir}
\author[a,b]{Ilangko Balasingham}
\affil[a]{Department of Electronic Systems, Norwegian University of Science and Technology, Trondheim, Norway}
\affil[b]{Intervention Centre, Oslo University Hospital, Oslo, Norway}

\authorinfo{Corresponding author: Mathias Ramm Haugland, E-mail: mathias.haugland@gmail.com}

\begin{document} 
\maketitle

\begin{abstract}
To cope with the growing prevalence of colorectal cancer (CRC), screening programs for polyp detection and removal have proven their usefulness. Colonoscopy is considered the best-performing procedure for CRC screening. To ease the examination, deep learning based methods for automatic polyp detection have been developed for conventional white-light imaging (WLI). Compared with WLI, narrow-band imaging (NBI) can improve polyp classification during colonoscopy but requires special equipment. We propose a CycleGAN-based framework to convert images captured with regular WLI to synthetic NBI (SNBI) as a pre-processing method for improving object detection on WLI when NBI is unavailable. This paper ﬁrst shows that better results for polyp detection can be achieved on NBI compared to a relatively similar dataset of WLI. Secondly, experimental results demonstrate that our proposed modality translation can achieve improved polyp detection on SNBI images generated from WLI compared to the original WLI. This is because our WLI-to-SNBI translation model can enhance the observation of polyp surface patterns in the generated SNBI images.
\end{abstract}


\vspace{-1em}
\section{INTRODUCTION}
\label{sec:intro}  
Colorectal cancer (CRC) is the third most incidental and the second deadliest cancer type globally \cite{sung2021}. It often develops from growths of glandular tissue known as polyps in the colon and rectum. In the early stage, polyps are benign and usually do not cause symptoms until some of them turn malignant. Early colon screening can have a great preventive effect. Colonoscopy is regarded as the best colon screening technique for CRC, where detection and removal of adenomatous polyps is the main goal. Colonoscopy is, however, a skill-dependent procedure relying heavily on the experience of endoscopists. Unfortunately, polyps might be missed or miss-classified in patients undergoing colonoscopy. This can lead to a late diagnosis of CRC and a poor prognosis. There have been technical efforts to reduce polyp miss rate and miss classification by improving the capability of colonoscopy devices. For example, narrow-band imaging (NBI)\cite{gono2015narrow}, a virtual chromoendoscopy technique, has been developed to enhance the observation of the scenes and visualization of the lesions. Compared to regular white-light imaging (WLI), which captures images in the full visible light spectrum, NBI uses an additional special filter to let the light of a specific narrow band of blue and green wavelengths pass through to the camera lens. The narrow-band light makes the thin capillary network and the thick blood vessels on the mucosal surface appear more distinctly, making it easier for the endoscopist to classify the polyps during  colonoscopy. To discern between benign polyps, adenomas, and cancer, the NICE (NBI International Colorectal Endoscopic)\cite{iwatate2020} classification is often used where NBI is available.

In recent years, many computer-aided diagnoses (CAD) systems have been proposed for automatic polyp detection \cite{hassan2021performance} as a solution for polyp miss rates. Among those, deep learning, in the form of convolutional neural networks (CNNs), has in recent years been the most prominent approach. Most of the proposed CNN-based automatic polyp detection algorithms are applied on WLI and the advantages of NBI over WLI have not been fully investigated and confirmed. CNNs also form the basis for generative adversarial networks (GANs), which are developed for generating real-looking synthetic data \cite{goodfellow2014}. GAN-based methods have many application areas, for instance, image-to-image translation \cite{zhu2017}. In this paper, we first apply the state-of-the-art object detection network EfficientDet-D0 \cite{tan2020} to confirm the advantages of NBI over WLI for automatic polyp detection. The initial results show that improved performance of polyp detection can be achieved with NBI. Therefore, we propose a GAN-based framework to convert colonoscopy images captured in WLI to synthetic NBI (SNBI) as a pre-processing method for the performance improvement of polyp detection on WLI when NBI is not available.

\section{MATERIALS AND METHODS}
\subsection{Method}
\label{sec:meth}
\begin{figure}[!t]
    \centering
    \includegraphics[scale=0.55]{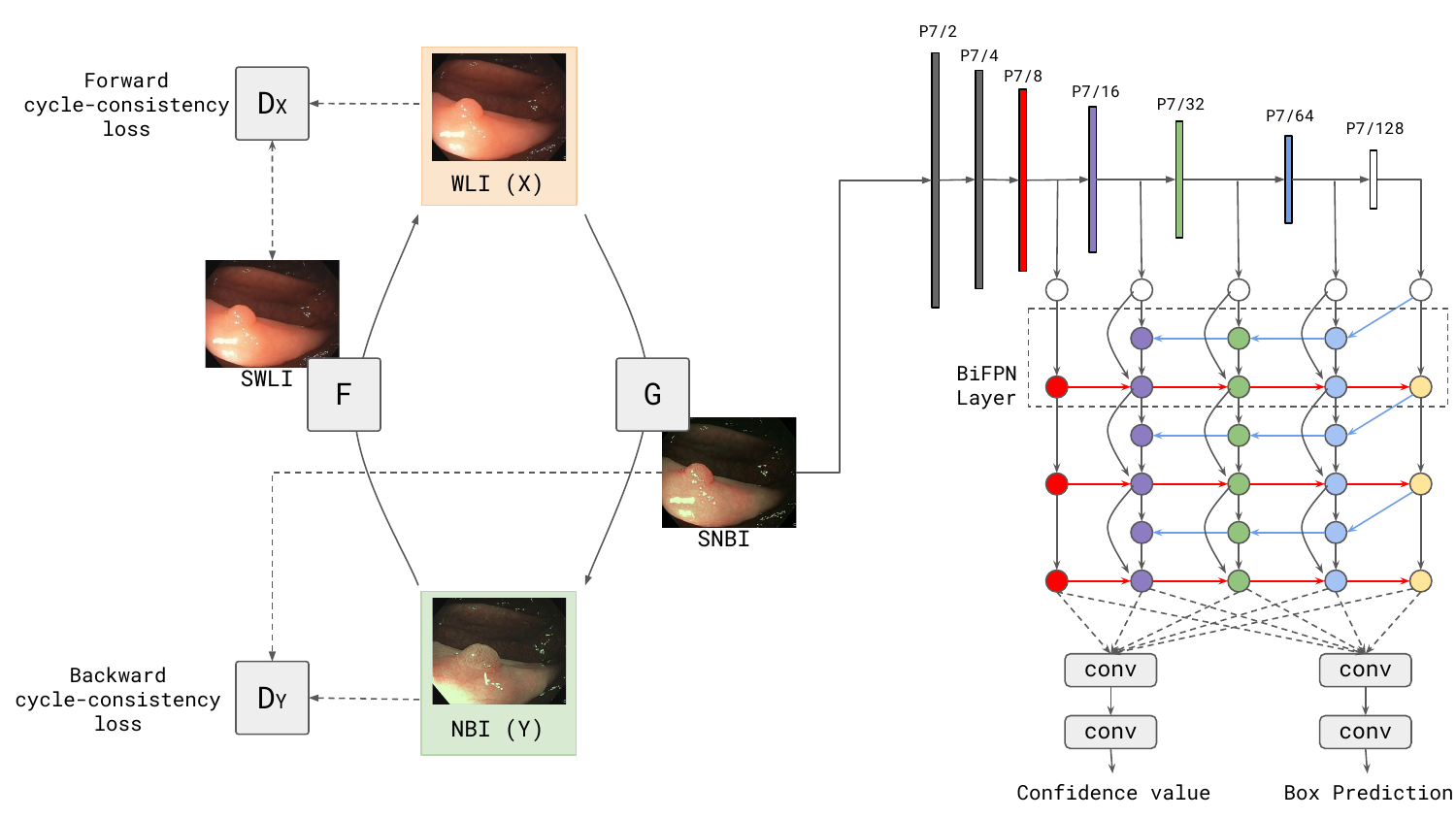}
    \vspace{-0.5em}
    \caption{The proposed framework. CycleGAN for modality translation (left), followed by EfficientDet-D0 for polyp detection (right).}
    \vspace{-1.0em}
    \label{fig:model}
\end{figure}

\vspace{-0.2em}
Figure \ref{fig:model} presents our proposed framework for the performance improvement of polyp detection in SNBI mode. The proposed framework consists of two networks: a network to perform an unpaired image-to-image translation and a detection network. For the image-to-image translation, we adapt CycleGAN \cite{zhu2017}, and for the detection model, we adapt EfficientDet-D0 \cite{tan2020}. The goal of CycleGAN is to learn patterns that connect two image domains, in this case WLI ($X$) and NBI ($Y$). CycleGAN is made up of two generators, \textit{G} and \textit{F}, and two discriminators, $D_X$ and $D_Y$. The aim is to learn the mapping $G: X \rightarrow Y$. For every training iteration, $G$ generates an SNBI image from a WLI image and $D_Y$ tries to discern between the generated SNBI image and a real NBI image. To ensure that the content of the SNBI image is not altered, \textit{F} converts each generated SNBI $\hat{y} = G(x)$ back to $X$. The difference between \textit{x} and $\hat{x} = F(\hat{y}) = F(G(x))$ is then calculated as what is called the forward cycle-consistency loss, which, along with the backward cycle-consistency loss and the two adversarial losses, makes up the loss function that trains both the generators and discriminators,

\vspace{-1em}
\begin{equation}
    \begin{aligned}
        \mathcal{L}(G,F,D_X,D_Y) &= \mathcal{L}_{GAN}(G,D_Y,X,Y) +
        \mathcal{L}_{GAN}(F,D_X,Y,X) +
        \lambda\mathcal{L}_{cyc}(G,F).
        \label{eq:cganloss}
    \end{aligned}
\end{equation}
\vspace{-1.5em}

$\mathcal{L}_{cyc}$ is the sum of the forward and the backward cycle-consistency losses, calculated as mean absolute error, and $\lambda$ is a weighting factor. During training, CycleGAN is given a WLI and a NBI image in each iteration. Because multiple datasets of different quality and origin are used for training and testing our models we propose a "semi-pairing" of the training data to improve model performance. This means that for each iteration, the two images are taken from the same video clip of the same colon.

We use EfficientDet-D0 to first evaluate the performance of polyp detection in the WLI modality against NBI, and second to assess the usefulness of the generated SNBI images for the improvement of automatic polyp detection. In an effort to have a fair modality comparison, the NBI and WLI datasets contain an equal number of similar images of the same polyps.

\subsection{Datasets}
\vspace{-0.2em}
\label{sec:dataset}
In this study, three public and one private dataset of still images and videos of hyperplastic (benign) and adenomatous polyps were used: 
\vspace{-0.9em}
\begin{enumerate}[label=\alph*)]
    \itemsep0em 
    \item PICCOLO \cite{sanchez_peralta2020}: This is a dataset of 76 lesions from 40 patients. The dataset contains both NBI and WLI of most polyps, NICE classification, and clinically annotated segmentation masks. All type 3 (deep submucosal invasive cancer) polyps were removed from the dataset. Then, all images were cropped on the sides, removing the black canvases and leaving the informative part of the image. Because both the EfficientDet-D0 and CycleGAN resize the images to 512x512, using more square-shaped images were believed to preserve the information in the images better. 
    \item OUS-NBI-ColonVDB: This is a private dataset containing 21 high-quality polyp videos: eleven hyperplastic and ten adenomatous polyps. The videos were recorded at Oslo University Hospital (OUS) in both NBI and WLI modes. The polyps are annotated by clinicians as binary segmentation masks in every frame. 
    \item KUMC \cite{li2021}: This is a mixed dataset of NBI and WLI frames obtained from 80 videos. Each frame is manually classified into adenoma or hyperplastic and annotated with a bounding box for each polyp present.
    \item Mesejo Videos \cite{mesejo2016}: This dataset comprises 76 polyp videos recorded in both NBI and WLI modes. The polyp distribution is as follows: 15 serrated, 21 hyperplastic, and 40 adenomatous. 
\end{enumerate}
The videos used were converted to still images and blurry images were manually removed. Binary segmentation masks were converted to bounding boxes.

\subsection{Evaluation metrics}
\label{sec:metrics}
\vspace{-0.2em}
For the evaluation of polyp detection performance, we derive recall and precision from the well-known medical terms true positive (TP), false positive (FP), and false negative (FN) as follows: 

\begin{equation}
    Precision = \frac{\sum_i TP_i}{\sum_i TP_i + \sum_i FP_i} = \frac{\sum_i TP_i}{all\ detections},
\end{equation}
\begin{equation}
    Recall = \frac{\sum_i TP_i}{\sum_i TP_i + \sum_i FN_i} = \frac{\sum_i TP_i}{all\ ground\ truths}.
\end{equation}
We identify a detection output as TP if the intersection over union (IoU) between the detected bounding box and the corresponding ground-truth bounding box is $>0.5$ and the confidence score is above a threshold. This threshold was found independently in each model based on the precision-recall curve from the validation data, keeping the trade-off between recall and precision.



\section{Results and Discussion}
\label{sec:res}
\subsection{Results on videos}
In this experiment, we trained our CycleGAN on the Mesejo dataset for 100 epochs with a constant learning rate of 0.0002. The detection model was trained three times: a) the WLI model was trained on the original WLI images of the PICCOLO dataset, b) the SNBI model was trained on the corresponding SNBI images obtained by our trained CycleGAN applied to the same WLI images, and c) the NBI model was trained on the original NBI images of the PICCOLO dataset. The NBI training set had an equal ammount of images of each lesion as the WLI/SNBI sets.

We split OUS-NBI-ColonVDB into a validation set (4 videos) and a test set (17 videos). To create as similar as possible datasets of both NBI and WLI videos, we removed several WLI and NBI frames so that there were equal numbers of similar video frames in both imaging modalities. Table \ref{tab:ex1} shows our results on the 17 test videos. 
\begin{table}[!h]
\centering
\caption{Results on the 17 test videos of OUS-NBI-ColonVDB.}
\label{tab:ex1}
\begin{tabular}{ll|l|l|l}
          & NBI   & WLI   & SNBI & SNBIx  \\ \hline
Precision & 0.757 & 0.55  & 0.634 & 0.457 \\
Recall    & 0.594 & 0.52  & 0.537 & 0.426 \\
F1        & 0.666 & 0.535 & 0.581 & 0.441
\end{tabular}
\end{table}

The results show that polyps can be detected more efficiently with NBI than with WLI. This is supported by the results from the SNBI images obtained from the WLI, which also show that the proposed transform improves polyp detection. The results of SNBIx were obtained when the NBI model was tested on the SNBI images. The SNBIx results are a good indicator that our CycleGAN generates SNBI images that resemble real NBI.

\subsection{Results on still images}
In this experiment, we first trained our CycleGAN on the Mesejo dataset for 100 epochs with a constant learning rate of 0.0002. The model was further trained on OUS-NBI-ColonVDB for 30 epochs with the same learning rate, followed by a linearly decaying learning rate for another 30 epochs. The detection model was trained on the PICCOLO dataset after being split into a training set (70\%) and a validation set (30\%). We trained two detection models: the WLI model was only trained on images in WLI mode, while the SNBI model was trained on the corresponding SNBI obtained by applying our trained CycleGAN on the same WLI images.  

We manually created our test set by handpicking images of 154 hyperplastic and 154 adenomatous polyps from the KUMC dataset. Table \ref{tab:ex2} shows the results obtained on our test images from the WLI and the SNBI models.
\begin{table}[!h]
    \centering
    \caption{Results on the KUMC dataset.}
    \label{tab:ex2}
    \begin{tabular}{lll}
                               & WLI   & SNBI  \\ \hline
    Precision (hyperplastic)   & 0.648 & \textbf{0.675}    \\
    Recall (hyperplastic)      & 0.669 & \textbf{0.677}    \\
    F1 (hyperplastic)          & 0.658 & \textbf{0.676}    \\ \hline
    Precision (adenomas)       & 0.734 & \textbf{0.742}    \\
    Recall (adenomas)          & 0.734 & \textbf{0.747}    \\
    F1 (adenomas)              & 0.734 & \textbf{0.744}          
    \end{tabular}
\end{table}

As seen in Table \ref{tab:ex2}, the SNBI model shows slightly better detection performance than the corresponding WLI model for both hyperplastic and adenomatous polyps. These results confirm that the detection model benefits from the enhanced SNBI polyps generated by our CycleGAN. 

\begin{figure}[!h]
    \centering
    \begin{tabular}{ccc}
         \includegraphics[scale=0.185]{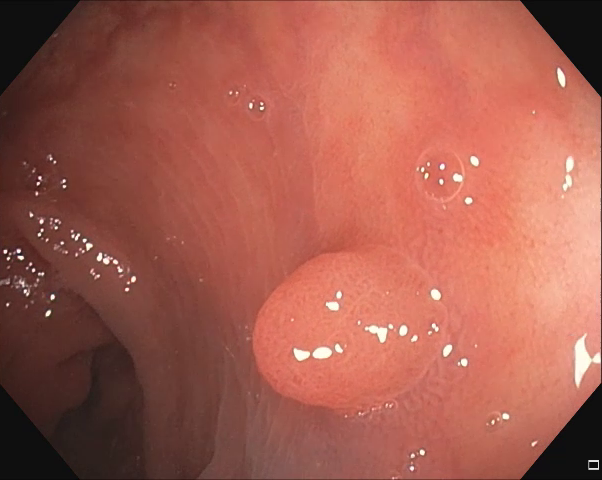} & \includegraphics[scale=0.185]{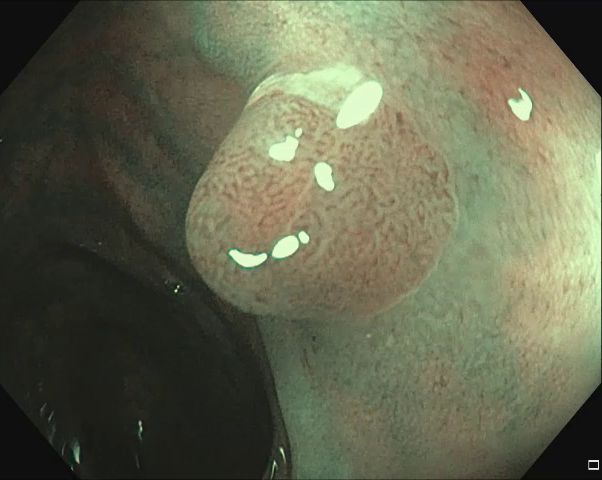} & \includegraphics[scale=0.185]{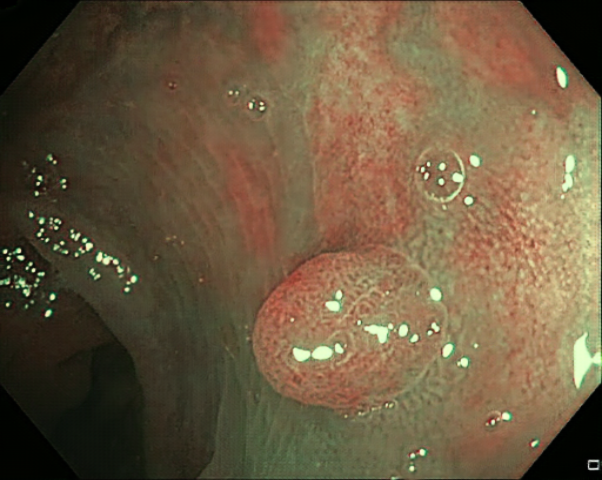} \\
        (a)  & (b)  & (c) 
    \end{tabular}
    \caption{The same polyp (from the PICCOLO set) is shown in (a) original WLI, (b) original NBI, and (c) SNBI generated from (a) by our CycleGAN.}
    \label{fig:WLI2SNBI}
    \vspace{0.5em}
\end{figure}

\begin{figure}[!h]
    \centering
    \begin{tabular}{ccc}
         \includegraphics[scale=0.26]{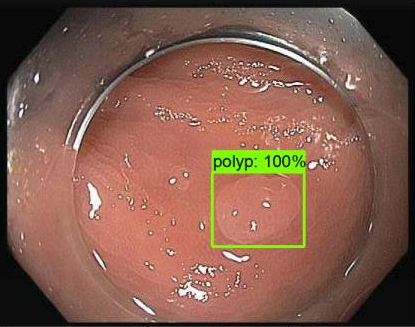} & \includegraphics[scale=0.26]{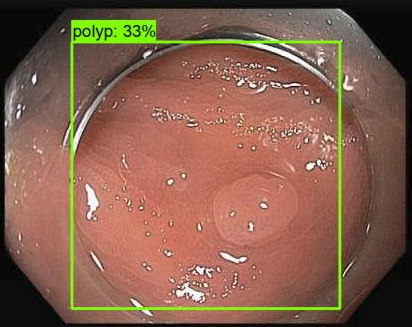} &  \includegraphics[scale=0.26]{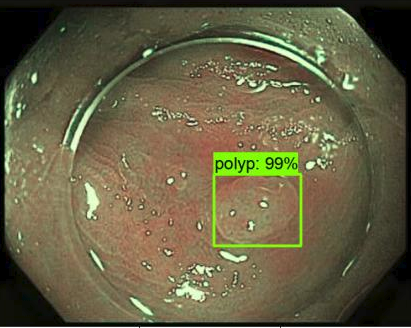} \\
         (a) Ground truth. & (b) Detection in WLI. & (c) Detection in SNBI.
    \end{tabular}
    \caption{An output detection shows that the detection model misses a hyperplastic polyp in WLI but detects it in SNBI.}
    \label{fig:Detection}
    \vspace{-0.75em}
\end{figure}

Figure \ref{fig:WLI2SNBI} shows an example of the WLI-to-SNBI conversion performed by our CycleGAN. Compared to the original WLI, the surface patterns on the SNBI can be better observed. Figure \ref{fig:Detection} demonstrates that our detection model misses the same polyp in the WLI, while it can detect it with a high confidence value and precise bounding box in the corresponding SNBI. This result clearly shows the benefits of our CycleGAN in generating synthetic NBI images for improved automatic polyp detection. 

Because colonoscopy equipment cannot capture NBI and WLI images from the same place at the same time, a fair comparison between the two modalities is challenging. Despite that our effort to compare the two neither was perfect, one can argue that NBI at least will be able to perform as well as the SNBI does. This is because NBI data has been used to create the SNBI transforms. The SNBI features that make polyps easier to detect than in the original WLI come arguably from NBI, implying the advantage of NBI over WLI in polyp detection.

\section{CONCLUSION AND FUTURE WORK}
\label{sec:con}
This paper has shown that better results for automatic polyp detection can be achieved on NBI compared to a relatively similar dataset of WLI. We, therefore, proposed a method based on unpaired image-to-image translation for converting colonoscopy images taken with WLI to SNBI. Our polyp detection results confirm that, compared to the original WLI, improved performance can be obtained on SNBI images. This shows the effectiveness of the proposed WLI-to-SNBI translation method and the advantage of using synthetic NBI images in automatic polyp detection. In our future work, we plan to extend the proposed method for automatic polyp classification based on the NICE classification. If the method can be confirmed, more polyps could be classified during colonoscopy. Thus, only the potentially dangerous polyps will be removed, reducing the time and labor-consuming biopsies and examinations under the microscope.

\vspace{-1em}
\acknowledgments 
This work is derived from the master's thesis \textit{Deep Learning for Polyp Detection from Narrow-Band Imaging} \cite{haugland2022} by Mathias Ramm Haugland, submitted in June 2022 at the Norwegian University of Science and Technology in Trondheim, Norway.

\vspace{-1em}
\bibliography{report} 
\bibliographystyle{spiebib} 

\end{document}